# Imprints of new physics in $\Lambda_b \to \Lambda^* l^+ l^-$ decay in non-universal $Z'$ model

**S. Biswas[1], S. Mahata[2], B. P. Nayak[3] and S. Sahoo[4]**

[1,2,4]Department of Physics, National Institute of Technology Durgapur
Durgapur-713209, West Bengal, India
[3]Department of Physics, Gandhi Institute for Technology
Gangapada, Bhubaneswar-752054, Odisha, India.

[1]E-mail: getswagata92@gmail.com, [4]E-mail: sukadevsahoo@yahoo.com

**Abstract**

Inspired by various updated tantalizing results of LHCb on baryonic sector we study the $\Lambda_b \to \Lambda^*(1520) l^+ l^-$ decay in non-universal $Z'$ model. We present the four-fold angular distributions of the decay in terms of transversity amplitudes. We structure the observables: differential branching ratio, lepton side forward backward asymmetries and polarization fractions in terms of the transversity amplitudes and incorporate the new physics (NP) terms in it. We reduce the number of the form factors using the improved Isgur-Wise relations. We study the observables in the standard model (SM) as well as in non-universal $Z'$ model. The predicted values are very interesting for the high energy physics community and results might provide prominent footprints of NP.

**Keywords:** Baryonic decays, $b$ hadrons, Flavour-changing neutral currents, Models beyond the standard model

**PACS numbers:** 14.70.Hp, 14.70.Pw, 13.30.Ce, 12.39.Hg

## I. Introduction

The loop induced rare flavour changing neutral current transitions containing $b$ hadrons are very famous over the last few decades for their high sensitive nature to the physics beyond the standard model (BSM). Although several predictions of the SM are in perfect agreement with many collider data up to now, there are many well-known anomalies: neutrino oscillations, unification of fundamental forces, baryon asymmetry, dark matter, strong CP violations, hierarchy problems etc., which cannot be explained by the SM and remain unsolved till now. One of the most powerful tools that can test lepton flavour universality (LFU) with the flavour changing neutral currents (FCNC) in the $b \to s l^+ l^-$ transition are $R_{K^{(*)}}$ which can be defined as: $R_{K^{(*)}} = \frac{Br(B \to K^{(*)} \mu^+ \mu^-)}{Br(B \to K^{(*)} e^+ e^-)}$. The LHCb has recorded the $R_K$ value recently as: $(R_K)_{new} =$

$0.846^{+0.060+0.016}_{-0.054-0.014}, 1 \leq q^2 \leq 6\,\text{GeV}^2$ [1] where $q^2$ is dilepton mass squared (in GeV$^2$). This result is deviated from the SM prediction $(R_K)_{SM} = 1.00 \pm 0.01$ [2] by $2.5\sigma$ discrepancy. Recently LHCb is also set the value of $R_{K^*}$ as [3]:

$$R_{K^*} = \begin{cases} 0.66^{0.11}_{-0.17} \pm 0.03, 0.045 \leq q^2 \leq 1.1 \\ 0.69^{0.11}_{-0.07} \pm 0.05, 1.1 \leq q^2 \leq 6.0 \end{cases}$$

The SM has predicted the $R_{K^*}$ values as:

$$R_{K^*}^{SM} = 0.906 \pm 0.028\,,$$
$$R_{K^*}^{SM} = 1.00 \pm 0.01$$

These experimental results are deviated from the predicted values of SM by $2.3\sigma$ and $2.5\sigma$ discrepancies respectively. In the reference [4] the values of $R_K$ and $R_{K^*}$ are experimented in Belle which are near to the SM value but with high uncertainties. The $R_D$ and $R_{D^*}$ observables provide the hints of LFU violation with the charged current transition $b \to cl\upsilon$. $R_{D^*}$ is experimented in Belle [5-7], BaBar [8] and LHCb [9]. The measured values of Belle [10] are:

$$R_D = 0.307 \pm 0.37 \pm 0.016\,,$$
$$R_{D^*} = 0.283 \pm 0.018 \pm 0.14\,.$$

These experimental results are higher than the SM values mentioned in ref. [11] and ref. [12] by $2.3\sigma$ and $3.4\sigma$ deviations respectively as below:

$$R_D^{SM} = 0.299 \pm 0.003\,,$$
$$R_{D^*}^{SM} = 0.258 \pm 0.005.$$

Some deviations are also incorporated to these lists nowadays which are: the branching ratios of $B \to K\mu^+\mu^-$ [13], $B \to K^*\mu^+\mu^-$ [14],$B \to \varphi\mu^+\mu^-$ [15] and various optimized observables in $B \to K^*\mu^+\mu^-$decay [16]. These deviations accelerate the exploration of NP in the B-physics world.

In this state of affairs, it is very interesting to establish NP in FCNC $b \to sl^+l^-$ transitions. Experimentalists can increase the size of data collection (currently performed by ATLAS, CMS, LHCb) and add new observables with different experimental domains (very soon begun by Belle II) [17]. On the other hand, theorists can modify the hadronic contributions on the decays in terms of local form factors and non-local charm-loop contributions [18]. Theoretically, it is always very interesting to study various hadronic decays induced by same quark level transitions. In fact, LHCb had provided many more fruitful results on not only mesonic decays but also baryonic decays which contain $b$ quarks. In recent days, the theoretical inspection of $\Lambda_b$ decays has received appreciable attention [19-27]. The LHCb searches for pentaquark states in $\Lambda_b \to \Lambda(\to pK^-)J/\psi$ which provide information about $\Lambda_b \to \Lambda(\to pK^-)l^+l^-$ decay [28]. According to [29], the dominant contribution is obtained from $\Lambda^* \equiv \Lambda(1520)$. Here $\Lambda^*$ has spin parity $J^P = 3/2^-$ and decays via $N\bar{K}$ pair under strong interaction. Due to these characteristics $\Lambda^*$ can be identifiable among the closely lying $\Lambda(1600)$, $\Lambda(1405)$ and $\Lambda(1116)$ ($\Lambda(1116)$ decays via $N\pi$ pair under weak interaction), all of which owning parity $J^P = 1/2^{\pm}$. Previously, Sahoo et al. [27] have studied various observables of $\Lambda_b \to \Lambda l^+l^-$ decays in non-universal $Z'$ model. In this model, the NP is allowed to contribute at tree level by $Z'$-mediated flavour changing $b \to q(q = s, d)$ decays where $Z'$ boson couples to the flavour-changing part $\bar{q}b$ as well as to the leptonic part $l^+l^-$ [30-34]. As the $Z'$ boson is not discovered yet the mass of $Z'$ boson is taken as arbitrary in various grand unified theories (GUTs). Recently LHC Drell-Yan data has announced as $M_{Z'} > 4.4$ TeV by Bandopadhyay et al., where the additional $U(1)$ coupling is similar to the $SU(2)_L$ coupling [35]. The upper bound of $Z'$ mass is set at 9 TeV from the study of the mass difference of $B_s$ meson in extended standard model [36].

To study the bounds in the NP couplings generated in our $Z'$ model we have used several observables of B meson decays. Mainly the bound on quark couplings are obtained from $B_s - \bar{B}_s$ mixing and the leptonic couplings are constrained from several exclusive and inclusive decays of B meson. Following the ref. [25, 37], we have reduced the number of form factors from14 to 2 using the heavy quark effective theory (HQET) on the basis of MCN quark model. The results obtained from MCN quark model in the reference [38] are not being followed very well by the soft collinear effective theory (SCET) assumptions.

This paper is organized as follows. In section II, we discuss the effective Hamiltonian of the decay. In section III, we present various tools of the decay which include the kinematics of the decay, hadronic and leptonic parts, reduction of the number of form factors using Isgur-Wise relations and structures of the observables. In section IV, the contribution of non-universal $Z'$ model is incorporated. The numerical analysis is presented in section V. Finally, we have discussed the consequences of our investigation in conclusion part of section VI.

## II. Effective Hamiltonian

We consider the effective Hamiltonian for $b \to sl^+l^-$ transition consisting the terms where the heavy degrees of freedom is integrated out in the short distance Wilson coefficients $C_i$ along with the set of operators $O_i$ describing the long distance physics as [25, 37],

$$\mathcal{H}^{eff} = -\frac{G_F}{2\sqrt{2}\pi} V_{tb} V_{ts}^* \alpha_e \sum_{r=7,9,10} C_i O_i, \tag{1}$$

The operators can be defined as,

$$\begin{aligned} O_7 &= \frac{m_b}{e} [\bar{s}\sigma^{\mu\nu}(1+\gamma_5)b] F_{\mu\nu} \\ O_9 &= [\bar{s}\gamma_\mu(1-\gamma_5)b][\bar{l}\gamma^\mu l] \\ O_{10} &= [\bar{s}\gamma_\mu(1-\gamma_5)b][\bar{l}\gamma^\mu\gamma_5 l], \end{aligned} \tag{2}$$

where $G_F$ is the Fermi coupling constant, $\alpha_e = e^2/4\pi$ is the fine structure constant and $V_{tb}V_{ts}^*$ are the CKM matrix elements. Among these operators $O_7$ is the electromagnetic operator and $O_9$ and $O_{10}$ are the semileptonic penguin operators. Here, all the three operators $O_{7,9,10}$ will contribute in the SM but only $O_{9,10}$ operators are sensitive to our non-universal $Z'$ model. Therefore, we can include the NP effects by modifying the Wilson coefficients $C_9$ and $C_{10}$. We have taken the $b$ quark mass as the running mass from the modified minimal subtraction scheme. The $V_{ub}V_{us}^*$ contribution is neglected as $V_{ub}V_{us}^* \ll V_{tb}V_{ts}^*$ and that is why CP violation is absent for this decay.

# III. The $\Lambda_b \to \Lambda^* l^+ l^-$ decay

## IIIA. Kinematics of decay

We consider the $\Lambda_b \to \Lambda^* l^+ l^-$ decay processes in terms of their momenta and spin as below [25, 37]:

$$\Lambda_b(p, s_{\Lambda_b}) \to \Lambda^*(k, s_{\Lambda^*}) l^+(q_1) l^-(q_2)\,,$$

$$\Lambda^*(k, s_{\Lambda^*}) \to N(k_1, s_N)\bar{K}(k_2)\,, \tag{3}$$

where $p, k, k_1, k_2, q_1$ and $q_2$ are the momenta of $\Lambda_b, \Lambda^*, N, \bar{K}$, the positively and negatively charged leptons respectively and $s_{\Lambda_b,\Lambda^*,N}$ are the projections of the spins of the baryons on to z-axis in their respective rest frames. The momentum for dilepton pair can be written as,

$$q^\mu = q_1^\mu + q_2^\mu\,. \tag{4}$$

From momentum conservation we can write as, $k^\mu = k_1^\mu + k_2^\mu$ and $p^\mu = k^\mu + q^\mu$. Our description of kinematics follows the convention of LHCb for this decay identified as $\theta_{\Lambda^*} = \theta_b$ and $\phi = \chi$ [39, 40]. Here $\theta_l$ is the angle made by $l^-$ with respect to the positive z axis in $l^+l^-$ rest frame, $\theta_{\Lambda^*}$ is defined as the angle made by the nucleon with the positive z axis in $N\bar{K}$ rest frame and $\phi$ is the angle formed between the decay planes of the dilepton pair and the hadron pair. The momenta can be defined in the dilepton rest frame as follows,

$$q_1^\mu|_{2l-RF} = (E_l, -|\mathrm{q}_{2l}|\sin\theta_l\,, 0, -|\mathrm{q}_{2l}|\cos\theta_l)\,,\; q_2^\mu|_{2l-RF} = (E_l, |\mathrm{q}_{2l}|\sin\theta_l\,, 0, |\mathrm{q}_{2l}|\cos\theta_l)\,,$$

where $|\mathrm{q}_{2l}| = \beta_l \frac{\sqrt{q^2}}{2}\,, \beta_l = \sqrt{1 - \frac{4m_l^2}{q^2}}$. Similarly other momenta in the $N\bar{K}$ rest frame, characterized by $k^2 = m_{\Lambda^*}^2$, can be defined as [37],

$$k_1^\mu|_{N\bar{K}-RF} = (E_N, |\mathrm{k}_{NK}|\sin\theta_{\Lambda^*}\cos\phi\,, |\mathrm{k}_{NK}|\sin\theta_{\Lambda^*}\sin\phi\,, |\mathrm{k}_{NK}|\cos\theta_{\Lambda^*})\,,$$

$$k_2^\mu|_{N\bar{K}-RF} = (E_K, -|\mathrm{k}_{NK}|\sin\theta_{\Lambda^*}\cos\phi\,, -|\mathrm{k}_{NK}|\sin\theta_{\Lambda^*}\sin\phi\,, -|\mathrm{k}_{NK}|\cos\theta_{\Lambda^*}),$$

where $E_N = \frac{k^2 + m_N^2 - m_{\bar{K}}^2}{2\sqrt{k^2}}\,, E_K = \frac{k^2 - m_N^2 + m_{\bar{K}}^2}{2\sqrt{k^2}}$, $|\mathrm{k}_{N\bar{K}}| = \frac{\sqrt{\lambda(k^2, m_N^2, m_{\bar{K}}^2)}}{2\sqrt{k^2}}$ and $\lambda(k^2, m_N^2, m_{\bar{K}}^2)$ is the triangular function (defined as, $\lambda(a, b, c) = a^2 + b^2 + c^2 - 2ab - 2bc - 2ac$).

## IIIB. Hadronic and leptonic helicity amplitudes of decay

Considering the parameterization between the hadronic and leptonic parts we can structure the matrix element of the four-body $\Lambda_b \to \Lambda^*(\to N\bar{K}) l^+ l^-$ as following [37]

$$\mathcal{M}(s_{\Lambda_b}, s_N, \lambda_1, \lambda_2) = \sum_{s_{\Lambda^*}} \mathcal{M}_{\Lambda_b}^{\lambda_1,\lambda_2}(s_{\Lambda_b}, s_{\Lambda^*}) \mathcal{M}_{\Lambda^*}(s_{\Lambda^*}, s_N)\,, \tag{5}$$

where $\mathcal{M}_{\Lambda^*}(s_{\Lambda^*}, s_N)$ represents the matrix element for $\Lambda^* \to N\bar{K}$ decay. This decay is discussed detail in the ref. [37]. The decay can be represented in terms of hadronic and leptonic helicity amplitudes as,

$$\mathcal{M}_{\Lambda_b}^{\lambda_1,\lambda_2}(s_{\Lambda_b}, s_{\Lambda^*}) = -\frac{G_F \alpha_e}{\sqrt{2}\pi} V_{tb} V_{ts}^* \sum_{h=L,R} \frac{1}{4} \left[ \sum_{\lambda} \eta_\lambda H_{VA,\lambda}^{h,s_{\Lambda_b},s_{\Lambda^*}} L_{h,\lambda}^{\lambda_1,\lambda_2} \right]. \tag{6}$$

The hadronic and leptonic helicity amplitudes are the projection of the full hadronic and leptonic amplitudes upon the direction of polarization of virtual gauge boson which decays in to the dilepton pair. And also $\eta_t = 1, \eta_\pm = -1$. According to the choice of gauge boson polarizations $\bar{\epsilon}_\mu(\lambda)$ for various polarization states $\lambda = 0, \pm 1, t$, the leptonic amplitudes can be written as [37]

$$L_{L(R),\lambda}^{\lambda_1,\lambda_2} = \bar{\epsilon}^\mu(\lambda) \langle \bar{l}(\lambda_1) l(\lambda_2) | \bar{l} \gamma_\mu (1 \mp \gamma_5) l | 0 \rangle . \tag{7}$$

The detailed expressions of the lepton helicity amplitudes are discussed in Appendix A. The hadronic helicity amplitudes can be represented in terms of the Wilson Coefficients and form factors which are defined in the ref. [37]. In our work, we are interested in the transversity amplitudes [24, 37, 41] $A_{\perp(\parallel)_1}^{L,R}, A_{\perp(\parallel)_0}^{L,R}$ and $A_{\perp(\parallel)t}^{L,R}$. Here, the tranversity amplitudes $B_{\perp(\parallel)_1}^{L,R}$ are zero. The transversity amplitudes are defined as the linear combinations of the helicity amplitudes. These expressions of transversity amplitudes are defined in Appendix B.

### IIIC. Reduction of form factors with improved Isgur-Wise relations

Following the ref. [37], we use the improved Isgur-Wise relations to reduce the number of form factors as a consequence of HQET spin symmetry. Here, we can consider two different kinematic scenarios: either for a soft departing $\Lambda^*$ baryon (in low recoil limit or high-$q^2$ region) or for an energetic $\Lambda^*$ baryon (in large recoil limit or low-$q^2$ region). In this work, we have used the numerical results presented in reference [38]. These results obey the HQET expectations very well but do not follow the SCET effects. We have assumed 10% uncorrelated uncertainties on the form factors in our calculation. With all these consideration and improved Isgur-Wise relations we have structured the form factors in terms of two leading Isgur-Wise form factors $\xi_{1,2}$ as [37],

$$f_\perp^{V,A} = C_0^{(v)} \frac{(\xi_1 \mp \xi_2)}{m_{\Lambda_b}}, \tag{8}$$

$$f_0^{V,A} = \left( C_0^{(v)} + \frac{C_1^{(v)} s_\pm}{2m_{\Lambda_b}(m_{\Lambda_b} \pm m_{\Lambda^*})} \right) \frac{\xi_1}{m_{\Lambda_b}} \mp \left( C_0^{(v)} - \frac{\left(2C_0^{(v)} + C_1^{(v)}\right) s_\pm}{2m_{\Lambda_b}(m_{\Lambda_b} \pm m_{\Lambda^*})} \right) \frac{\xi_2}{m_{\Lambda_b}}, \tag{9}$$

$$f_\perp^{T(5)} = C_0^{(t)} \left( \frac{(\xi_1 \mp \xi_2)}{m_{\Lambda_b}} \pm \frac{s_\pm}{m_{\Lambda_b}(m_{\Lambda_b} \pm m_{\Lambda^*})} \frac{\xi_2}{m_{\Lambda_b}} \right), \tag{10}$$

$$f_0^{T(5)} = C_0^{(t)} \frac{(\xi_1 \mp \xi_2)}{m_{\Lambda_b}}, \tag{11}$$

$$f_t^V(q^2) = \left( C_0^{(v)} + C_1^{(v)} \left( 1 - \frac{s_-}{2m_{\Lambda_b}(m_{\Lambda_b} - m_{\Lambda^*})} \right) \right) \frac{\xi_1}{m_{\Lambda_b}}$$
$$+ \left( C_0^{(v)} \left( 1 - \frac{s_-}{m_{\Lambda_b}(m_{\Lambda_b} - m_{\Lambda^*})} \right) \right.$$
$$\left. + C_1^{(v)} \left( 1 - \frac{s_-}{2m_{\Lambda_b}(m_{\Lambda_b} - m_{\Lambda^*})} \right) \right) \frac{\xi_2}{m_{\Lambda_b}}, \quad (12)$$

$$f_t^A(q^2) = \left( C_0^{(v)} + C_1^{(v)} \left( 1 - \frac{s_+}{2m_{\Lambda_b}(m_{\Lambda_b} + m_{\Lambda^*})} \right) \right) \frac{\xi_1}{m_{\Lambda_b}}$$
$$- \left( C_0^{(v)} \left( 1 - \frac{s_+}{m_{\Lambda_b}(m_{\Lambda_b} + m_{\Lambda^*})} \right) \right.$$
$$\left. + C_1^{(v)} \left( 1 - \frac{s_+}{2m_{\Lambda_b}(m_{\Lambda_b} + m_{\Lambda^*})} \right) \right) \frac{\xi_2}{m_{\Lambda_b}}, \quad (13)$$

$$f_g^V = f_g^A = f_g^T = f_g^{T5} = 0 . \quad (14)$$

Where,

$$C_0^{(v)} = 1 - \frac{\alpha_s C_F}{4\pi} \left( 3ln\left(\frac{\mu}{m_b}\right) + 4 \right) + \mathcal{O}(\alpha_s^2) ,$$
$$C_1^{(v)} = \frac{\alpha_s C_F}{2\pi} + \mathcal{O}(\alpha_s^2) ,$$
$$C_0^{(t)} = 1 - \frac{\alpha_s C_F}{4\pi} \left( 5ln\left(\frac{\mu}{m_b}\right) + 4 \right) + \mathcal{O}(\alpha_s^2) , \quad (15)$$
$$s_\pm = \left( m_{\Lambda_b} \pm m_{\Lambda^*} \right)^2 - q^2 . \quad (16)$$

These form factors are used in the expressions of the transversity amplitudes. Here $\alpha_s$ is the strong coupling constant and $C_F$ is the quadratic casimir in the fundamental representation of the gauge group.

**IIID. Angular distributions**

Proceeding with all these calculations, we have found the expression of four fold angular distribution for this decay as [25, 37],

$$\frac{d^4\mathcal{B}}{dq^2 d\cos\theta_l\, d\cos\theta_{\Lambda^*}\, d\phi}$$
$$= \frac{3}{8\pi}\Big((K_{1c}\cos\theta_l + K_{1cc}cos^2\theta_l + K_{1ss}sin^2\theta_l)cos^2\theta_{\Lambda^*}$$
$$+ (K_{2c}\cos\theta_l + K_{2cc}cos^2\theta_l + K_{2ss}sin^2\theta_l)sin^2\theta_{\Lambda^*}$$
$$+ (K_{3ss}sin^2\theta_l)sin^2\theta_{\Lambda^*}\cos\phi + (K_{4ss}sin^2\theta_l)sin^2\theta_{\Lambda^*}\, sin\,\phi\cos\phi$$
$$+ (K_{5s}\, sin\,\theta_l + K_{5sc}\, sin\,\theta_l\cos\theta_l)\, sin\,\theta_{\Lambda^*}\cos\theta_{\Lambda^*}\cos\phi$$
$$+ (K_{6s}\, sin\,\theta_l + K_{6sc}\, sin\,\theta_l\cos\theta_l)\, sin\,\theta_{\Lambda^*}\cos\theta_{\Lambda^*}\, sin\,\phi\Big)\,. \qquad (17)$$

Here $K_{\{1c,\ldots,6sc\}}$ are the angular coefficients which are expressed in terms of transversity amplitudes. The detail is described in the Appendix C. We have kept the masses of the final states and considered the mass corrections of the order $\mathcal{O}\big(m_l/\sqrt{q^2}\big)$ and $\mathcal{O}(m_l^2/q^2)$.

## IIIE. Observables

From angular integrals of the four fold angular distribution of eq. (11) for this decay we have calculated various observables as the combinations of the angular co-efficients as [25, 37]

$$\mathcal{O}(\omega) = \int \frac{d^4\mathcal{B}}{dq^2 d\cos\theta_l\, d\cos\theta_{\Lambda^*}\, d\phi}\,\omega(q^2,\theta_l,\theta_{\Lambda^*},\phi)d\cos\theta_l\, d\cos\theta_{\Lambda^*}\, d\phi\,. \qquad (18)$$

The observables are structured as follows:

Taking $\omega = 1$ and all the considerations for the transversity amplitudes we obtain the expression of differential branching ratio in terms of angular coefficients (described in Appendix C) as

$$\frac{d\mathcal{B}}{dq^2} = 2K_{1ss} + K_{1cc} + 2K_{2cc}\,. \qquad (19)$$

Another powerful tool for searching new physics in rare B decays is the forward-backward asymmetry which is given by,

$$A_{FB}(q^2) = \frac{\int_0^1 \frac{d^2\mathcal{B}}{dq^2 d\cos\theta_l} d\cos\theta_l - \int_{-1}^0 \frac{d^2\mathcal{B}}{dq^2 d\cos\theta_l} d\cos\theta_l}{\int_0^1 \frac{d^2\mathcal{B}}{dq^2 d\cos\theta_l} d\cos\theta_l + \int_{-1}^0 \frac{d^2\mathcal{B}}{dq^2 d\cos\theta_l} d\cos\theta_l}\,. \qquad (20)$$

Considering $\omega = sgn[\cos\theta_l]/(d\mathcal{B}/dq^2)$ we obtain the expression of lepton side forward-backward asymmetry as

$$A_{FB}(q^2) = \frac{3}{2}\frac{(K_{1c} + 2K_{2c})}{\big(K_{1cc} + 2(K_{1ss} + K_{2cc} + 2K_{2ss})\big)}\,. \qquad (21)$$

Another interesting observable is the longitudinal polarization fraction for the dilepton system as

$$F_L = 1 - \frac{2(K_{1cc} + 2K_{2cc})}{\big(K_{1cc} + 2(K_{1ss} + K_{2cc} + 2K_{2ss})\big)}\,, \qquad (22)$$

with the consideration of $\omega = (2 - 5cos^2\theta_l)/(d\mathcal{B}/dq^2)$. For the $\Lambda_b \to \Lambda(\to N\pi)l^+l^-$ decay hadron side forward-backward asymmetry and mixed forward-backward asymmetry can be calculated with the respective weight factors as $\omega = sgn[\cos\theta_{\Lambda^*}]/(d\mathcal{B}/dq^2)$ and $\omega = sgn[\cos\theta_{\Lambda^*}\cos\theta_l]/(d\mathcal{B}/dq^2)$. But here $\Lambda^*$ decays strongly so these observables become zero.

## IV. Contribution of $Z'$ boson in $\Lambda_b \to \Lambda^* l^+ l^-$ decays

Our non-universal $Z'$ model is one of the most economical models in which NP introduces a heavy new gauge boson $Z'$ with its quark couplings and leptonic couplings. Here an extra $U(1)'$ gauge group is incorporated with the SM group [32, 42] and the FCNC transitions are successfully explained to the tree level by the characteristic of the $Z'$couplings with fermions. To construct theory for different families we have considered the family non-universal $Z'$ couplings in BSM theories. The leptonic ouplings of $Z'$ are different for all the three lepton generations was first predicted by Chaudhuri, Hockney and Lykken [43-45]. Here, our assumption is that our $Z'$ couplings are diagonal and non-universal in nature.

Including the NP part the $U(1)'$ currents can be represented as,

$$J_\mu = \sum_{i,j} \bar{\psi}_j \gamma_\mu \left[\epsilon_{\psi_{L_{ij}}} P_L + \epsilon_{\psi_{R_{ij}}} P_R\right] \psi_i. \quad (23)$$

Here, the sum is done over all fermions $\psi_{i,j}$ and the chiral couplings of the gauge boson are denoted by $\epsilon_{\psi_{R,L_{ij}}}$. Usually FCNCs originate at the tree level in both the LH and RH sectors. So we can write the matrices of chiral $Z'$ couplings in the fermion mass eigenstate basis as, $B_{ij}^{\psi_L} \equiv \left(V_L^{\psi} \epsilon_{\psi_L} V_L^{\psi^\dagger}\right)_{ij}$ , $B_{ij}^{\psi_R} \equiv \left(V_R^{\psi} \epsilon_{\psi_R} V_R^{\psi^\dagger}\right)_{ij}$. The $Z'\bar{b}s$ couplings can be generated in the form as below

$$\mathcal{L}_{FCNC}^{Z'} = -g'\left(B_{sb}^L \bar{s}_L \gamma_\mu b_L + B_{sb}^R \bar{s}_R \gamma_\mu b_R\right) Z'^\mu + h.c.. \quad (24)$$

Here $g'$ is the gauge coupling correspond to the extra $U(1)'$ group and the effective Hamiltonian can be structured as,

$$H_{eff}^{Z'} = \frac{8G_F}{\sqrt{2}} \left(\rho_{sb}^L \bar{s}_L \gamma_\mu b_L + \rho_{sb}^R \bar{s}_R \gamma_\mu b_R\right)\left(\rho_{ll}^L \bar{l}_L \gamma_\mu l_L + \rho_{ll}^R \bar{l}_R \gamma_\mu l_R\right), \quad (25)$$

where $\rho_{ll}^{L,R} \equiv \frac{g' M_Z}{g M_{Z'}} B_{ll}^{L,R}$ , $g'$ and $g$ are the gauge couplings of $Z'$ and $Z$ bosons (where $g = \frac{e}{\sin\theta_W \cos\theta_W}$) respectively. Here, we need to simplify with some considerations: (i) we have ignored the kinetic mixing throughout the whole analysis as it represents the redefinition of unknown couplings, (ii) we have also ignored $Z - Z'$ mixing [32, 46-49] angle because it is constrained as less than $10^{-3}$ by Bandyopadhyay et al. [35] and recently at LHC it was found as of the order $10^{-4}$ by Bobovnikov et al. [50], (iii) we have avoided the crucial effects of renormalization group (RG) evolution between the W boson mass ($M_W$) and the $Z'$ mass ($M_{Z'}$) scales, (iv) to neglect too many free parameters we consider that the exceptional contribution in the flavour changing $b - s - Z'$ part is offered by the flavour-off-diagonal left-handed couplings of quarks [51-55] only. Accompanying all these simplifications mentioned previously the following parameters have constrained recently with the LHC Drell-Yan data: the $Z'$ mass ($M_{Z'}$), the $Z-Z'$ mixing angle and the extra $U(1)'$ gauge coupling. Bandyopadhyay et al. have constrained the $Z'$ mass as $M_{Z'} < 4.4$ TeV by [35]. The value of $\left|\frac{g'}{g}\right|$ is not estimated till now. But we can presume the value of $\left|\frac{g'}{g}\right|$ as unity because both $U(1)$ and $U(1)'$ gauge groups generate from the same genesis of grand unified theory and $\left|\frac{M_Z}{M_{Z'}}\right| \sim 0.1$ for TeV-scale $Z'$. The existence of $Z'$ boson with the same fermionic couplings similar to the SM $Z$ boson is suggested by all four experiments of LEP. If we consider $|\rho_{sb}^L| \sim |V_{tb}V_{ts}^*|$, then the order of $B_{sb}^L$ will be of $10^{-3}$. Including all the non-universal $Z'$ couplings the effective Hamiltonian for $b \to sl^+l^-$ becomes

$$H_{eff}^{Z'} = -\frac{2G_F}{\sqrt{2}\pi} V_{tb}V_{ts}^* \left[\frac{B_{sb}^L B_{ll}^L}{V_{tb}V_{ts}^*} (\bar{s}b)_{V-A}(\bar{l}l)_{V-A} - \frac{B_{sb}^L B_{ll}^R}{V_{tb}V_{ts}^*} (\bar{s}b)_{V-A}(\bar{l}l)_{V+A}\right] + h.c., \quad (26)$$

where $B_{sb}^L$ is the left-handed coupling of $Z'$ boson with quarks, $B_{ll}^L$ and $B_{ll}^R$ are the left-handed and right-handed couplings with the leptons respectively. The coupling parameter consists of a NP weak phase term which is related as, $B_{sb}^L = |B_{sb}^L| e^{-i\varphi_s^l}$.

According to the discussion in Sec II we have included the NP terms with the assistance of eq. (1) and eq. (2) as follows [56]:

$$C_9^{NP} = \frac{4\pi B_{sb}^L}{\alpha V_{tb}V_{ts}^*}(B_{ll}^L + B_{ll}^R),$$

$$C_{10}^{NP} = \frac{4\pi B_{sb}^L}{\alpha V_{tb}V_{ts}^*}(B_{ll}^L - B_{ll}^R). \quad (27)$$

With all above considerations we have studied the observables defined in previous section with our NP approach.

## V. Numerical Analysis

In this paper, we have studied the observables: differential branching ratio, lepton side forward-backward asymmetry and longitudinal polarization fraction and examine its sensitivity to the SM as well as to the NP. We have observed the variations of these observables in the whole kinematic region for three leptonic channels in the context of non-universal $Z'$ model. We have adapted the Wilson coefficients $C_9$ and $C_{10}$ with NP terms as mentioned in eq. (27) and eq. (B7). To structure the $C_9^{eff}$ term we have included the short distance effects [56] in our work. Traditionally with the short distance contributions there are some long-distance effects too due to the $c\bar{c}$ intermediate states. According to Briet-Wigner approximation this long- distance effects can be included by modifying $C_9^{eff}$ with the term,

$$Y_{res} = \frac{3\pi}{\alpha^2}\kappa(3C_1 + C_2 + 3C_3 + C_4 + 3C_5 + C_6)\sum_{V_n=\psi(ns)} \frac{\Gamma(V_n \to l^+l^-) m_{V_n}}{m_{V_n}^2 - \hat{s}m_b^2 - i m_{V_n}\Gamma_{V_n}} \quad . \quad (28)$$

Here the factor $\kappa$ is introduced to include the known factorizable and the unknown nonfactorizable effects for the recent experimental data of $B \to \psi(nS)X_s$ decays. But the approximation carries a problem also; it counts the partonic and hadronic degrees of freedom doubly. In this situation [57], we need to avoid the double-counting in long-distance effects. Therefore we can consider the KS approach. According to this approach the effect is calculated using the factorizable part only and considering the dispersion relation with the experimental value of $R_c(s) = \sigma(e^+e^- \to c\bar{c})/\sigma(e^+e^- \to \mu^+\mu^-)$. Now we have to calculate the nonfactorizable part to continue with the experimental data of $B \to \psi(\psi')X_s$ decays. According to the detailed discussions of the reference [58], a harsh disaster of quark-hadron duality in the resonance region is offered due to the $c\bar{c}$ contribution to $B \to X_s l^+l^-$. This occurs because of the integration over whole phase space of the absolute square of the correlator. On the other hand, from the recent study [59] of $B \to Kl^+l^-$ it has been inspected that duality violations from higher resonances in high $q^2$ region spoil the particularity of theoretical predictions for (partially) integrated branching ratios of $B \to Kl^+l^-$ at the level of some percentages. Therefore, we can expect that the effects of charmonium resonances will not alter our conclusion much. In our work we have ignored the resonance effects and focus on the short distance effects to incorporate the NP contributions.

Here, it is essential to fix the NP coupling parameters from various low energy experiments. The $b - s - Z'$ coupling $|B_{sb}^L|$ and the NP weak phase $\varphi_s^l$ are constrained from $B_s - \bar{B}_s$ mixing data of UTfit Collaboration [60-66]. The left-handed and right-handed leptonic coupling terms are constrained by various exclusive and inclusive decays. The values of NP couplings are recollected in Table-1 [27, 56].

**Table-1: Numerical values of coupling parameters**

| Scenarios | $\lvert B_{sb}\rvert \times 10^{-3}$ | $\varphi_s^l$ (in degree) | $S_{ll} \times 10^{-2}$ | $D_{ll} \times 10^{-2}$ |
|---|---|---|---|---|
| $S_1$ | $(1.09 \pm 0.22)$ | $(-72 \pm 7)°$ | $(-2.8 \pm 3.9)$ | $(-6.7 \pm 2.6)$ |
| $S_2$ | $(2.20 \pm 0.15)$ | $(-82 \pm 4)°$ | $(-1.2 \pm 1.4)$ | $(-2.5 \pm 0.9)$ |

To enhance the NP effects we have put in the maximum values of the coupling terms and selected two sets of values of the parameters as mentioned below.

**Scenario 1:** According to the range of $S_1$ in Table-1 the maximum magnitude of the NP coupling parameter $|B_{sb}|$ and NP weak phase angle $\varphi_s^l$ are set as $|B_{sb}| = (1.31 \times 10^{-3})$ and $\varphi_s^l = -65°$. And the lepton couplings are considered as $S_{ll} = (1.1 \times 10^{-2})$ and $D_{ll} = (-4.1 \times 10^{-2})$.

**Scenario 2:** Similarly with the limit for $S_2$ in Table-1 the enhanced effect of NP couplings are set as $|B_{sb}| = (2.35 \times 10^{-3})$ and $\varphi_s^l = -78°$. And the leptonic couplings are taken as $S_{ll} = (1.2 \times 10^{-2})$ and $D_{ll} = (-1.6 \times 10^{-2})$.

Utilizing all these numerical data from Table-1 and Appendix D we have started to explore the $\Lambda_b \to \Lambda^* l^+l^-$ decays in a background of NP. Actually the pictorial illustration can elucidate the responsiveness of NP model on the decays distinctly. Therefore we have plotted the above mentioned observables for all the three channels of $\Lambda_b \rightarrow \Lambda^* l^+ l^-$ decays in whole kinematic region of $q^2$. In Fig. 1a, Fig. 1b and Fig. 1c, we have varied the differential branching ratio for muonic, electronic and taunic channels respectively. The differential branching ratio is high at low-$q^2$ region and drops with the increment of $q^2$ in both the scenarios. The blue line indicates the SM value whereas the violet and red line indicates the 1st and 2nd scenario values respectively. The differential branching ratio has maximum value for scenario 2 which indicates the sensitivity of NP on $\Lambda_b \rightarrow \Lambda^* l^+ l^-$ decays. The value of the differential branching ratio for the electronic channel and taunic channel is two orders lesser and one order greater than the muonic one respectively. Here, another thing one can observe that the value for 1st scenario is closer to the SM one and it is more deviated from the SM in 2nd scenario. Incorporating the uncertainties coming from the input values and form factors, the maximum and minimum branching ratio values are calculated for three different kinematic regions in Table-2.

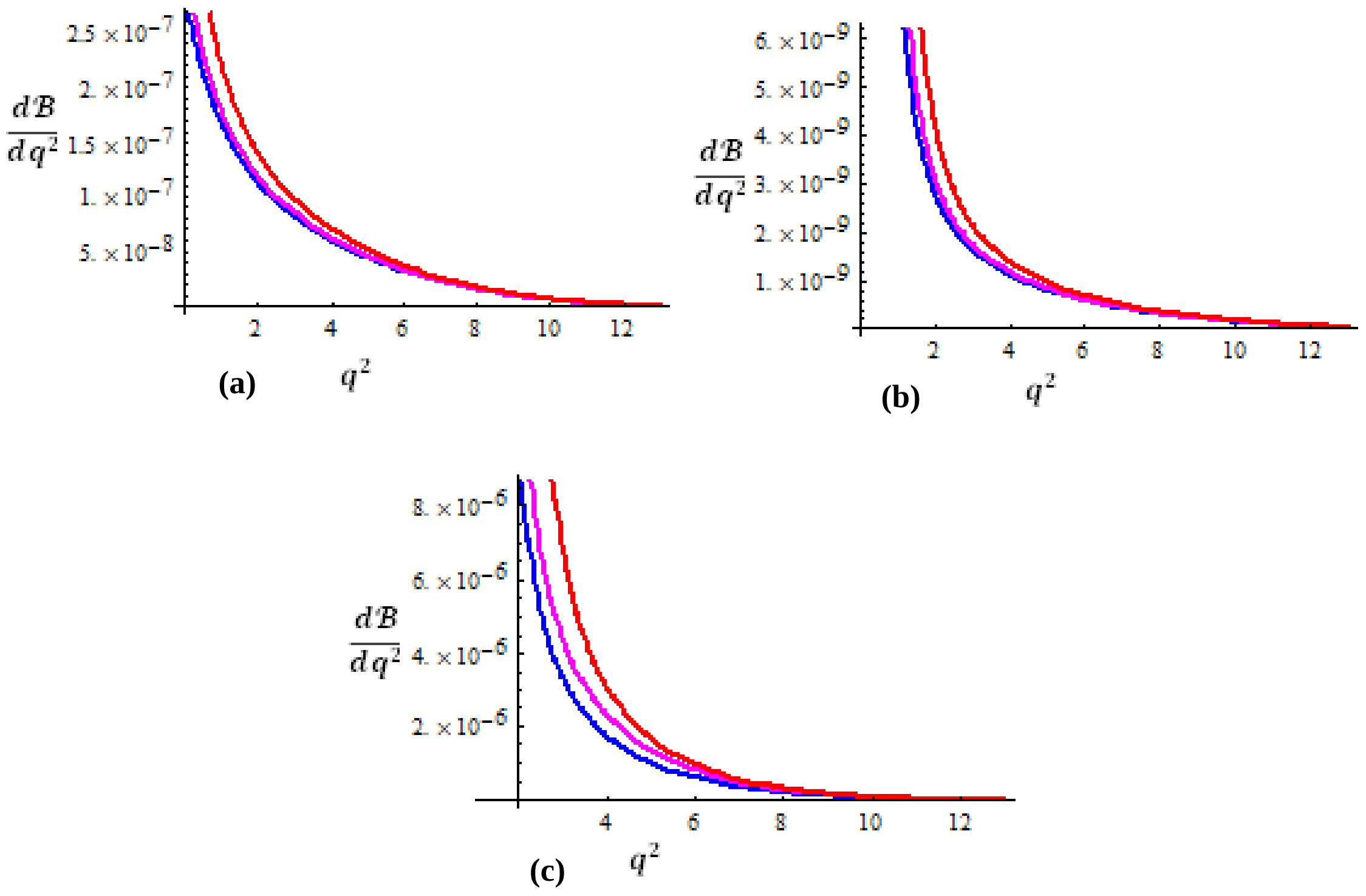

**Fig. 1: Variation of differential branching ratio for $\Lambda_b \rightarrow \Lambda^* l^+ l^-$ within allowed kinematic region of $q^2$ in SM and in NP for (a) muonic, (b) electronic and (c) taunic channel.**

**Table-2: Predicted values of branching ratios for $\Lambda_b \rightarrow \Lambda^* l^+ l^-$ decay over entire $q^2$ phase space in SM, 1st and 2nd scenarios**

| Decay | | SM | 1st Scenario | 2nd Scenario |
|---|---|---|---|---|
| $\Lambda_b \rightarrow \Lambda^* \mu^+ \mu^-$ | Max. value | $7.116 \times 10^{-7}$ | $7.344 \times 10^{-7}$ | $8.363 \times 10^{-7}$ |
| | Min. value | $4.828 \times 10^{-7}$ | $4.976 \times 10^{-7}$ | $5.643 \times 10^{-7}$ |
| $\Lambda_b \rightarrow \Lambda^* e^+ e^-$ | Max. value | $7.142 \times 10^{-9}$ | $7.372 \times 10^{-9}$ | $8.389 \times 10^{-9}$ |
| | Min. value | $5.409 \times 10^{-9}$ | $5.697 \times 10^{-9}$ | $6.811 \times 10^{-9}$ |
| $\Lambda_b \rightarrow \Lambda^* \tau^+ \tau^-$ | Max. value | $1.387 \times 10^{-5}$ | $1.824 \times 10^{-5}$ | $2.871 \times 10^{-5}$ |
| | Min. value | $9.268 \times 10^{-6}$ | $1.224 \times 10^{-5}$ | $1.878 \times 10^{-5}$ |

We have plotted the variation of forward-backward asymmetries ($A_{FB}$) in Fig. 2a, Fig. 2b and Fig. 2c for muonic, electronic and taunic channels respectively. The description of the coloured lines is discussed previously. It is very interesting accessory of high energy physics to investigate NP. The nature of variations of $A_{FB}$ with $q^2$ is of similar type for muonic and electronic channels which are plotted in Fig. 2a and Fig. 2b. The $A_{FB}$ values are positive in high recoil limit and it is negative for rest of the kinematic region. The increment of the value from the SM confirms the sensitivity of NP on it and it reaches maximum for 2nd scenario. The zero crossings are indicated in Fig. 3a and Fig. 3b for $\Lambda_b \rightarrow \Lambda^* \mu^+ \mu^-$ and $\Lambda_b \rightarrow \Lambda^* e^+ e^-$ decay respectively. Here, we observe that the zero crossing is shifting towards origin for the electronic channel. Among the three scenarios the 2nd scenario has the least zero crossing value. It indicates that the higher value of NP couplings flip the nature of $A_{FB}$ faster. This result agrees with our previous work [67]. Another very exciting thing we observe that the zero crossing is absent for tau channel, it has negative value throughout the whole kinematic region. Actually the zero crossing point is the particular value of $q^2$ in which the uncertainties arising due to the form factors are cancelled out [68]. This cancellation in the dominant hadronic uncertainties leads the observable to a less theoretical uncertainty. But for the tau channel these uncertainties may affect the $A_{FB}$ in some extent. The maximum and minimum values of $A_{FB}$ over the allowed entire phase space for three channels are presented in Table-3.

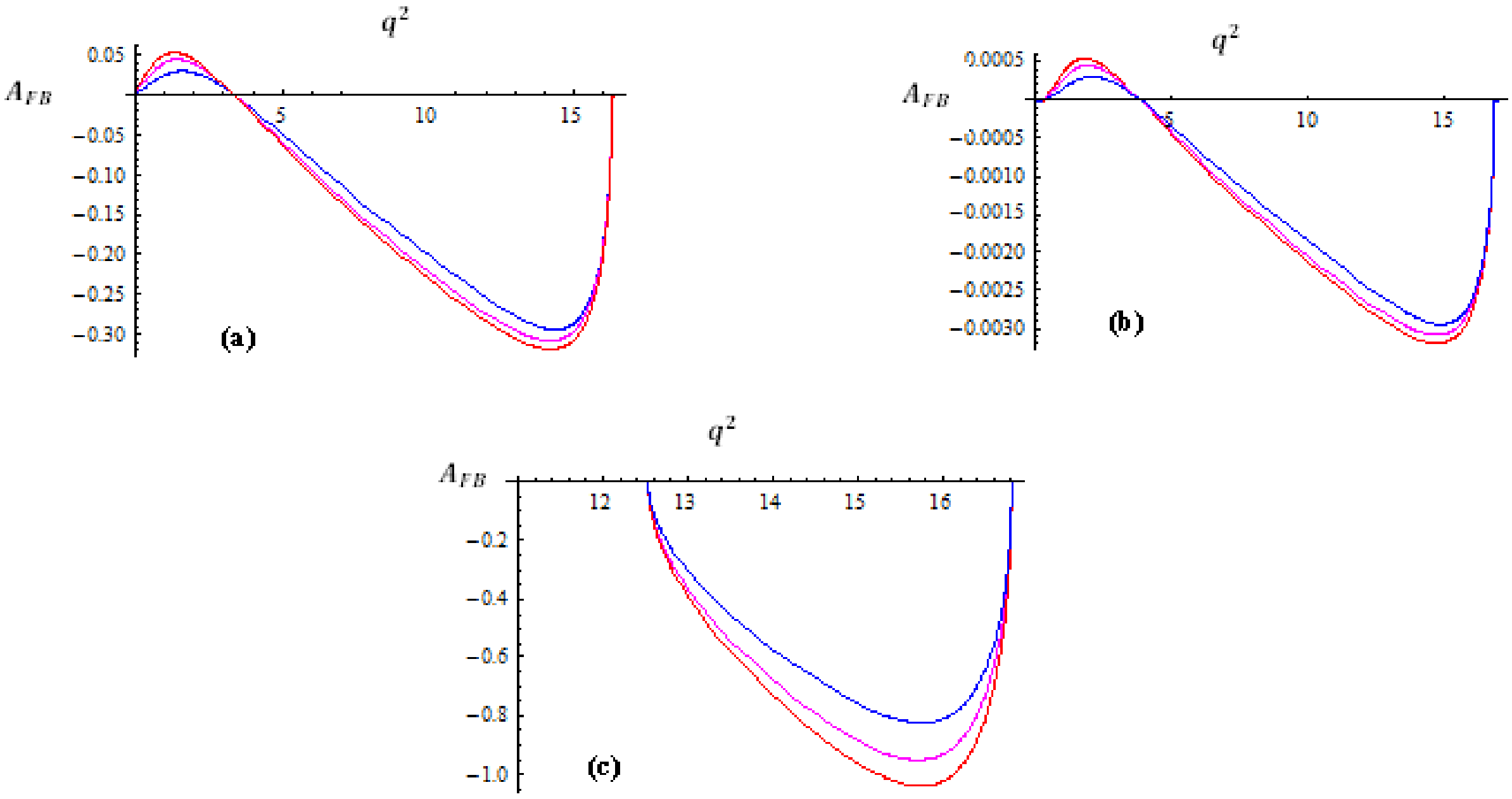

**Fig. 2: Variation of differential branching ratio for $\Lambda_b \to \Lambda^* l^+ l^-$ within allowed kinematic region of $q^2$ in SM and in NP for (a) muonic, (b) electronic and (c) taunic channel.**

**Table-3: Predicted average values of forward backward asymmetries for $\Lambda_b \to \Lambda^* l^+ l^-$ decay over entire $q^2$ phase space in SM, 1st and 2nd scenarios**

| Decay | | SM | 1st Scenario | 2nd Scenario |
|---|---|---|---|---|
| $\Lambda_b \to \Lambda^* \mu^+ \mu^-$ | Max. value | −0.089 | −0.082 | −0.079 |
| | Min. value | −0.135 | −0.128 | −0.105 |
| $\Lambda_b \to \Lambda^* e^+ e^-$ | Max. value | −0.00075 | −0.00066 | −0.00060 |
| | Min. value | −0.00105 | −0.00095 | −0.00090 |
| $\Lambda_b \to \Lambda^* \tau^+ \tau^-$ | Max. value | −0.717 | −0.680 | −0.615 |
| | Min. value | −0.854 | −0.806 | −0.785 |

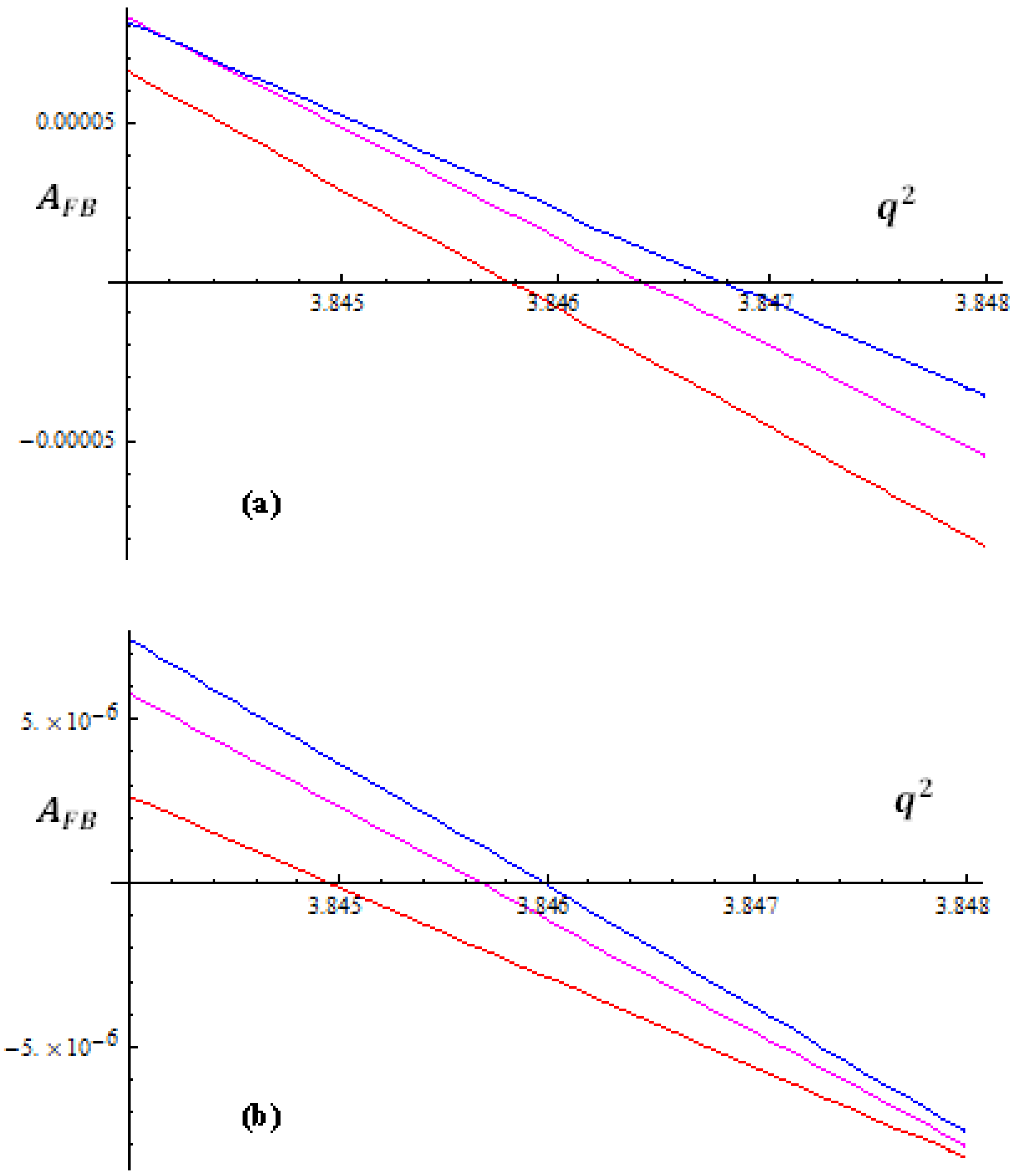

**Fig. 3: Variation of lepton side forward backward asymmetry for (a) $\Lambda_b \to \Lambda^* \mu^+ \mu^-$ and (b) $\Lambda_b \to \Lambda^* e^+ e^-$ in low $q^2$ region to locate the position of zero crossing**

Let us consider the longitudinal polarization fraction for the dilepton system. In the advance 2000's this observable had drawn attention, named as the well-known Polarization Puzzle. We have varied the polarization fraction $(F_L)$ for the total allowed kinematic region in Fig. 4a, Fig. 4b and Fig. 4c for muonic, electronic and taunic channels respectively. Similar to the observable $A_{FB}$ the variation due to electronic and muonic channels is of same type. Within the allowed $q^2$ region, the values of $F_L$ is non-zero which lies between 0.96 to 0.44 for muonic channel and 0.094 to 0.05 for electronic channel. Here, we observe a kink in SM line for muonic channel in low $q^2$ region but it is absent for electronic channel. The variation of tau channel is completely different from the other two. Within the allowed $q^2$ region for the taunic channel the $F_L$ value restricts between 0.53 to 0.435. The maximum and minimum $F_L$ values are calculated in Table-4 for all three channels.

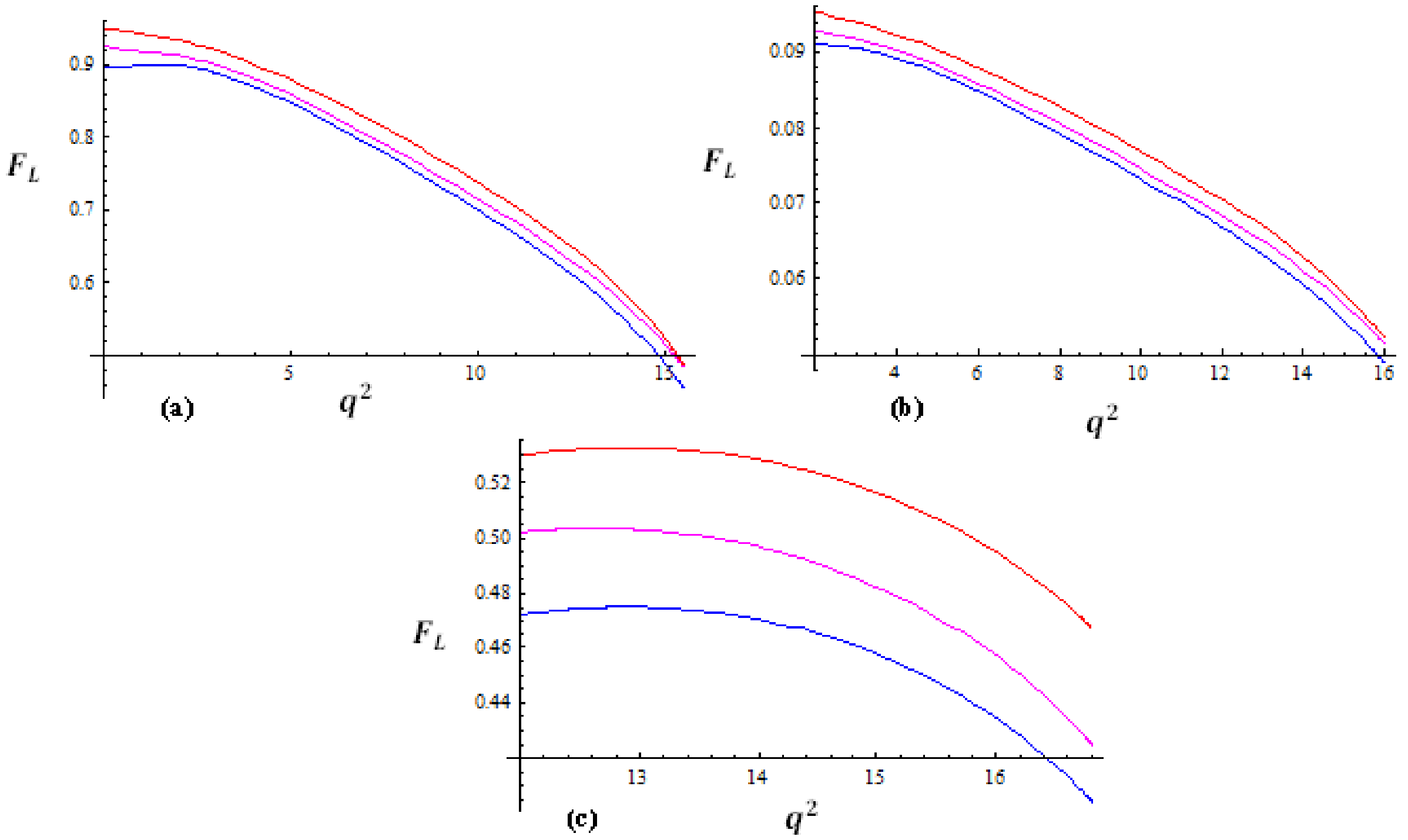

**Fig. 4: Variation of polarization fraction for $\Lambda_b \to \Lambda^* l^+ l^-$ within allowed kinematic region of $q^2$ in SM and in NP for (a) muonic, (b) electronic and (c) taunic channel.**

**Table-4: Predicted values of polarization fraction for $\Lambda_b \to \Lambda^* l^+ l^-$ decay over entire $q^2$ phase space in SM, 1st and 2nd scenarios**

| Decay | | SM | 1st Scenario | 2nd Scenario |
|---|---|---|---|---|
| $\Lambda_b \to \Lambda^* \mu^+ \mu^-$ | Max. value | 0.778 | 0.794 | 0.816 |
| | Min. value | 0.605 | 0.621 | 0.685 |
| $\Lambda_b \to \Lambda^* e^+ e^-$ | Max. value | 0.069 | 0.080 | 0.089 |
| | Min. value | 0.048 | 0.050 | 0.058 |
| $\Lambda_b \to \Lambda^* \tau^+ \tau^-$ | Max. value | 0.455 | 0.481 | 0.514 |
| | Min. value | 0.354 | 0.356 | 0.435 |

## VI. Conclusion

To the best of our knowledge, the LHCb has studied the $\Lambda_b \to \Lambda^* l^+ l^-$ decay for muonic channel and found $\Lambda^*(1520)$ is decaying into a pair of a proton and a negatively charged kaon $(pK^-)$ [28, 69]. They have reproduced $\Lambda^*(1520)$ but the full angular analysis of the decay is yet to be studied experimentally. In this work, we have studied some interesting characteristics of $\Lambda_b \to \Lambda^* l^+ l^-$ decays in the SM as well as in the context of the non-universal Z′ model. According to the hadronic elements fourteen form factors are involved in the decay. But here we have used the improved relations of Isgur-Wise form factors to reduce the number of form factors and also the uncertainties originated from them. We have investigated the variation of differential branching ratio over the allowed kinematic region in Fig. 1. The nature of the variability of differential branching ratio for all three leptonic channels are quiet similar where SM and scenario 1 curves are very close to each other and scenario 2 has larger branching ratio. The divergence of branching ratio values with NP scenarios in Table 2 shows the sensitivity of the NP in the decay.

The variation of forward backward asymmetries within allowed kinematic region is shown in Fig. 2 and the zero crossing points are indicated in Fig. 3. We can see that the zeroes are crossed at large recoil for both muonic and electronic channels. The positions of zero crossing points depend on the considered scenarios and this shows the responsiveness of NP in the observable. In ref. [37], the zero crossings are obtained at low recoil region whereas in the ref. [25], the observable crossed the zero at low $q^2$ regime (which is in agreement with our investigation). For experimentalists the measurement of polarization fraction is very interesting as it explains the polarization puzzle more satisfactorily with the help of BSM physics which is in contrary to the early predictions established on helicity arguments, which predict the observable as 1. According to our inspection the polarization fraction is nearly 1 for muonic channel and 0.1 for electronic channel whereas for taunic channel it is regulated within 0.53 to 0.435.

Considering the uncertainties arriving from the input values of Appendix D and form factors we have calculated the maximum and minimum values of the observables in Table-2, 3, 4. It is very interesting and trust worthy to test and confirm the NP with $b \to s l^+ l^-$ transition. Therefore we have explored NP in baryonic sector with this $\Lambda^*(1520)$ baryon. It would be also very exciting to attain this decay experimentally. We can expect that the predicted values of the branching ratios, forward backward asymmetries and the polarization fractions would help out the experimental community to access it in near future.

**Acknowledgement**
We thank the reviewer for suggesting valuable improvements of our manuscript. Biswas and Mahata thank NIT Durgapur for providing fellowship for their research. Sahoo is grateful to the SERB, DST, Govt. of India for financial support through project (EMR/2015/000817).

**Appendix A:**

The leptonic helicity amplitudes can be written explicitly as,

$$\bar{u}_{l_1}\gamma_\mu(1 \mp \gamma_5)v_{l_2}, \bar{\epsilon}^\mu(\lambda)\bar{u}_{l_1}\gamma_\mu(1 \mp \gamma_5)v_{l_2}, \qquad (A1)$$

Here we assume that the lepton $l_1$ is the negatively charged $l^-$ and $l_2$ is the positively charged $l^+$. From the ref [70] the explicit expressions of the spinor for lepton $l_1$ is expressed as,

$$\bar{u}_{l_1}(\lambda) = \begin{bmatrix} \sqrt{E_l + m_l}\chi^u_\lambda \\ 2\lambda\sqrt{E_l - m_l}\chi^u_\lambda \end{bmatrix}, \chi^u_{+\frac{1}{2}} = \begin{bmatrix} \cos\frac{\theta_l}{2} \\ \sin\frac{\theta_l}{2} \end{bmatrix}, \chi^u_{-\frac{1}{2}} = \begin{bmatrix} -\sin\frac{\theta_l}{2} \\ \cos\frac{\theta_l}{2} \end{bmatrix}. \quad (A2)$$

Another spinor for the lepton $l_2$ which is moving opposite to lepton $l_1$,

$$v_{l_2}(\lambda) = \begin{bmatrix} \sqrt{E_l - m_l}\chi^v_{-\lambda} \\ -2\lambda\sqrt{E_l + m_l}\chi^v_{-\lambda} \end{bmatrix}, \chi^v_{+\frac{1}{2}} = \begin{bmatrix} \sin\frac{\theta_l}{2} \\ -\cos\frac{\theta_l}{2} \end{bmatrix}, \chi^v_{-\frac{1}{2}} = \begin{bmatrix} \cos\frac{\theta_l}{2} \\ \sin\frac{\theta_l}{2} \end{bmatrix}. \quad (A3)$$

From the ref [70] we have studied the two component spinors which are related as $\chi^v_{-\lambda} = \xi_\lambda \chi^u_\lambda$ and $\xi_\lambda = 2\lambda e^{-2i\lambda\varphi}$ where $\varphi$ is the azimuthal angle.

Along with all these considerations we have structured the expressions of the lepton helicity amplitudes which are collected below,

$$L_L^{+\frac{1}{2}+\frac{1}{2}} = -L_R^{-\frac{1}{2}-\frac{1}{2}} = \sqrt{q^2}\,(1+\beta_l)\,, \quad (A4)$$

$$L_L^{-\frac{1}{2}-\frac{1}{2}} = -L_R^{+\frac{1}{2}+\frac{1}{2}} = \sqrt{q^2}\,(1-\beta_l)\,, \quad (A5)$$

$$L_{L,+1}^{-\frac{1}{2}-\frac{1}{2}} = L_{R,+1}^{-\frac{1}{2}-\frac{1}{2}} = L_{L,-1}^{+\frac{1}{2}+\frac{1}{2}} = L_{R,-1}^{+\frac{1}{2}+\frac{1}{2}} = \sqrt{2}m_l\sin\theta_l, \quad (A6)$$

$$L_{L,-1}^{-\frac{1}{2}-\frac{1}{2}} = L_{R,-1}^{-\frac{1}{2}-\frac{1}{2}} = L_{L,+1}^{+\frac{1}{2}+\frac{1}{2}} = L_{R,+1}^{+\frac{1}{2}+\frac{1}{2}} = -\sqrt{2}m_l\sin\theta_l\,, \quad (A7)$$

$$L_{L,-1}^{+\frac{1}{2}-\frac{1}{2}} = -L_{R,+1}^{-\frac{1}{2}+\frac{1}{2}} = -\sqrt{\frac{q^2}{2}(1-\beta_l)(1-\cos\theta_l)}\,, \quad (A8)$$

$$L_{L,-1}^{-\frac{1}{2}+\frac{1}{2}} = -L_{R,+1}^{+\frac{1}{2}-\frac{1}{2}} = \sqrt{\frac{q^2}{2}(1+\beta_l)(1+\cos\theta_l)}\,, \quad (A9)$$

$$L_{R,-1}^{+\frac{1}{2}-\frac{1}{2}} = -L_{L,+1}^{-\frac{1}{2}+\frac{1}{2}} = -\sqrt{\frac{q^2}{2}(1+\beta_l)(1-\cos\theta_l)}\,, \quad (A10)$$

$$L_{R,-1}^{-\frac{1}{2}+\frac{1}{2}} = -L_{L,+1}^{+\frac{1}{2}-\frac{1}{2}} = \sqrt{\frac{q^2}{2}(1-\beta_l)(1+\cos\theta_l)}\,, \quad (A11)$$

$$L_{L,0}^{+\frac{1}{2}+\frac{1}{2}} = -L_{L,0}^{-\frac{1}{2}-\frac{1}{2}} = L_{R,0}^{+\frac{1}{2}+\frac{1}{2}} = -L_{R,0}^{-\frac{1}{2}-\frac{1}{2}} = 2m_l\cos\theta_l, \quad (A12)$$

$$L_{L,0}^{+\frac{1}{2}-\frac{1}{2}} = L_{R,0}^{-\frac{1}{2}+\frac{1}{2}} = -\sqrt{q^2}\,(1-\beta_l)\sin\theta_l\,, \quad (A13)$$

$$L_{R,0}^{+\frac{1}{2}-\frac{1}{2}} = L_{L,0}^{-\frac{1}{2}+\frac{1}{2}} = -\sqrt{q^2}\,(1+\beta_l)\sin\theta_l\,, \quad (A14)$$

$$L_{L,t}^{+\frac{1}{2}+\frac{1}{2}} = L_{L,t}^{-\frac{1}{2}-\frac{1}{2}} = -L_{R,t}^{+\frac{1}{2}+\frac{1}{2}} = -L_{R,t}^{-\frac{1}{2}-\frac{1}{2}} = 2m_l\,, \quad (A15)$$

**Appendix B:**

$$A_{\perp 0}^{L,(R)} = -\sqrt{2}N\left(f_0^V \frac{(m_{\Lambda_b}+m_{\Lambda^*})}{\sqrt{q^2}}\frac{s_-\sqrt{s_+}}{\sqrt{6}m_{\Lambda^*}}C_+^{L,R} + \frac{2m_b}{q^2}f_0^T\sqrt{q^2}\frac{s_-\sqrt{s_+}}{\sqrt{6}m_{\Lambda^*}}C_7\right), \qquad (B1)$$

$$A_{\parallel 0}^{L,(R)} = \sqrt{2}N\left(f_0^A \frac{(m_{\Lambda_b}-m_{\Lambda^*})}{\sqrt{q^2}}\frac{s_+\sqrt{s_-}}{\sqrt{6}m_{\Lambda^*}}C_-^{L,R} + \frac{2m_b}{q^2}f_0^{T5}\sqrt{q^2}\frac{s_+\sqrt{s_-}}{\sqrt{6}m_{\Lambda^*}}C_7\right), \qquad (B2)$$

$$A_{\perp 1}^{L,(R)} = -\sqrt{2}N\left(f_\perp^V \frac{s_-\sqrt{s_+}}{\sqrt{3}m_{\Lambda^*}}C_+^{L,R} + \frac{2m_b}{q^2}f_\perp^T(m_{\Lambda_b}+m_{\Lambda^*})\frac{s_-\sqrt{s_+}}{\sqrt{3}m_{\Lambda^*}}C_7\right), \qquad (B3)$$

$$A_{\parallel 1}^{L,(R)} = -\sqrt{2}N\left(f_\perp^A \frac{s_+\sqrt{s_-}}{\sqrt{3}m_{\Lambda^*}}C_-^{L,R} + \frac{2m_b}{q^2}f_\perp^{T5}(m_{\Lambda_b}-m_{\Lambda^*})\frac{s_+\sqrt{s_-}}{\sqrt{3}m_{\Lambda^*}}C_7\right), \qquad (B4)$$

$$A_{\parallel t}^{L,(R)} = \pm\sqrt{2}N\left(f_t^A \frac{(m_{\Lambda_b}+m_{\Lambda^*})}{\sqrt{q^2}}\frac{s_-\sqrt{s_+}}{\sqrt{6}m_{\Lambda^*}}C'_{10}\right), \qquad (B5)$$

$$A_{\perp t} = \mp\sqrt{2}N\left(f_t^V \frac{(m_{\Lambda_b}-m_{\Lambda^*})}{\sqrt{q^2}}\frac{s_+\sqrt{s_-}}{\sqrt{6}m_{\Lambda^*}}C'_{10}\right). \qquad (B6)$$

Here the triangular function defined as, $\lambda(a,b,c) = a^2+b^2+c^2-2(ab+bc+ac)$ and $s_\pm = \left\{(m_{\Lambda_b}+m_{\Lambda^*})^2 - q^2\right\}$. The Wilson coefficients are structured as

$$C_+^{L,R} = \left(C_9^{eff} \mp C_{10}\right) + \left(C_9^{NP} \mp C_{10}^{NP}\right),$$
$$C_-^{L,R} = \left(C_9^{eff} \mp C_{10}\right) - \left(C_9^{NP} \mp C_{10}^{NP}\right). \qquad (B7)$$

The NP terms $C_9^{NP}$ and $C_{10}^{NP}$ are set to zero for SM analysis. The normalisation term is absorbed in the transversity amplitudes and can be defined as,

$$N = G_F\alpha_e V_{tb}V_{ts}^*\sqrt{\tau_{\Lambda_b}q^2\frac{\sqrt{\lambda(m_{\Lambda_b}^2, m_{\Lambda^*}^2, q^2)}}{3.\,2^{11}m_{\Lambda_b}^3\pi^5}\beta_l\mathcal{B}_{\Lambda^*}}\,, \beta_l = \sqrt{1-\frac{4m_l^2}{q^2}}\,.$$

Here $\mathcal{B}_{\Lambda^*}$ is defined as the branching ratio for $\Lambda^* \to N\bar{K}$ decay and $\tau_{\Lambda_b}$ is the $\Lambda_b$ lifetime.

**Appendix C:**

The expressions of angular coefficients can be written as below.

$$K_{1c} = -2\beta_l\left(Re\left(A_{\perp 1}^L A_{\parallel 1}^{*L}\right) - \{L\leftrightarrow R\}\right), \qquad (C1)$$

$$\begin{aligned}K_{1cc} = {} & \left(\left|A_{\parallel 1}^L\right|^2 + |A_{\perp 1}^L|^2 + \{L\leftrightarrow R\}\right) \\ & + \frac{2m_l^2}{q^2}\Big[\Big(\left|A_{\parallel 0}^L\right|^2 + |A_{\perp 0}^L|^2 - \left|A_{\parallel 1}^L\right|^2 - |A_{\perp 1}^L|^2 + |A_{\perp t}^L|^2 + \left|A_{\parallel t}^L\right|^2 + Re\left(A_{\parallel 0}^R A_{\parallel 0}^{*L}\right) \\ & + Re\left(A_{\parallel 1}^R A_{\parallel 1}^{*L}\right) + Re(A_{\perp 0}^R A_{\perp 0}^{*L}) + Re(A_{\perp 1}^R A_{\perp 1}^{*L}) - Re\left(A_{\parallel t}^R A_{\parallel t}^{*L}\right) - Re(A_{\perp t}^R A_{\perp t}^{*L}) \\ & + \{L\leftrightarrow R\}\Big)\Big], \end{aligned} \qquad (C2)$$

$$\begin{aligned}K_{1ss} = {} & \frac{1}{2}\left(2\left|A_{\parallel 0}^L\right|^2 + \left|A_{\parallel 1}^L\right|^2 + 2|A_{\perp 0}^L|^2 + |A_{\perp 1}^L|^2 + \{L\leftrightarrow R\}\right) \\ & + \frac{2m_l^2}{q^2}\Big[\Big(|A_{\perp t}|^2 + \left|A_{\parallel t}\right|^2 - \left|A_{\parallel 0}^L\right|^2 - |A_{\perp 0}^L|^2 + Re\left(A_{\parallel 0}^R A_{\parallel 0}^{*L}\right) + Re\left(A_{\parallel 1}^R A_{\parallel 1}^{*L}\right) \\ & + Re(A_{\perp 0}^R A_{\perp 0}^{*L}) + Re(A_{\perp 1}^R A_{\perp 1}^{*L}) - Re\left(A_{\parallel t}^R A_{\parallel t}^{*L}\right) - Re(A_{\perp t}^R A_{\perp t}^{*L}) \\ & + \{L\leftrightarrow R\}\Big)\Big], \end{aligned} \qquad (C3)$$

$$K_{2c} = -\frac{1}{2}\beta_l\left(Re\left(A_{\perp 1}^{L}A_{\parallel 1}^{*L}\right) + 3Re\left(B_{\perp 1}^{L}B_{\parallel 1}^{*L}\right) - \{L \leftrightarrow R\}\right), \tag{C4}$$

$$\begin{aligned} K_{2cc} = \frac{1}{4}&\left(\left|A_{\parallel 1}^{L}\right|^2 + |A_{\perp 1}^{L}|^2 + 3\left|B_{\parallel 1}^{L}\right|^2 + 3|B_{\perp 1}^{L}|^2 + \{L \leftrightarrow R\}\right) \\ &+ \frac{1}{2}\frac{m_l^2}{q^2}\Big[\Big(\left|A_{\parallel 0}^{L}\right|^2 + |A_{\perp 0}^{L}|^2 - \left|A_{\parallel 1}^{L}\right|^2 - |A_{\perp 1}^{L}|^2 + |A_{\perp t}^{L}|^2 + \left|A_{\parallel t}^{L}\right|^2 - 3\left|B_{\parallel 1}^{L}\right|^2 \\ &- 3|B_{\perp 1}^{L}|^2 + Re\left(A_{\parallel 0}^{R}A_{\parallel 0}^{*L}\right) + Re\left(A_{\parallel 1}^{R}A_{\parallel 1}^{*L}\right) + Re(A_{\perp 0}^{R}A_{\perp 0}^{*L}) + Re(A_{\perp 1}^{R}A_{\perp 1}^{*L}) \\ &- Re\left(A_{\parallel t}^{R}A_{\parallel t}^{*L}\right) - Re(A_{\perp t}^{R}A_{\perp t}^{*L}) + 3Re\left(B_{\parallel 1}^{R}B_{\parallel 1}^{*L}\right) + 3Re(B_{\perp 1}^{R}B_{\perp 1}^{*L}) \\ &+ \{L \leftrightarrow R\}\Big)\Big], \end{aligned} \tag{C5}$$

$$\begin{aligned} K_{2ss} = \frac{1}{8}&\left(2\left|A_{\parallel 0}^{L}\right|^2 + \left|A_{\parallel 1}^{L}\right|^2 + 2|A_{\perp 0}^{L}|^2 + |A_{\perp 1}^{L}|^2 + 3\left|B_{\parallel 1}^{L}\right|^2 + 3|B_{\perp 1}^{L}|^2 + \{L \leftrightarrow R\}\right) \\ &+ \frac{1}{2}\frac{m_l^2}{q^2}\Big[\Big(|A_{\perp t}|^2 + \left|A_{\parallel t}\right|^2 - \left|A_{\parallel 0}^{L}\right|^2 - |A_{\perp 0}^{L}|^2 + Re\left(A_{\parallel 0}^{R}A_{\parallel 0}^{*L}\right) + Re\left(A_{\parallel 1}^{R}A_{\parallel 1}^{*L}\right) \\ &+ Re(A_{\perp 0}^{R}A_{\perp 0}^{*L}) + Re(A_{\perp 1}^{R}A_{\perp 1}^{*L}) - Re\left(A_{\parallel t}^{R}A_{\parallel t}^{*L}\right) - Re(A_{\perp t}^{R}A_{\perp t}^{*L}) \\ &+ 2\sqrt{3}Re\left(B_{\parallel 1}^{L}A_{\parallel 1}^{*L}\right) - 2\sqrt{3}Re(B_{\perp 1}^{L}A_{\perp 1}^{*L}) + 3Re\left(B_{\parallel 1}^{R}B_{\parallel 1}^{*L}\right) + 3Re(B_{\perp 1}^{R}B_{\perp 1}^{*L}) \\ &+ \{L \leftrightarrow R\}\Big)\Big], \end{aligned} \tag{C6}$$

$$\begin{aligned} K_{3ss} = \frac{\sqrt{3}}{2}&\left(Re\left(B_{\parallel 1}^{L}A_{\parallel 1}^{*L}\right) - Re(B_{\perp 1}^{L}A_{\perp 1}^{*L}) + \{L \leftrightarrow R\}\right) \\ &- 2\sqrt{3}\frac{m_l^2}{q^2}\left(Re\left(B_{\parallel 1}^{L}A_{\parallel 1}^{*L}\right) - (B_{\perp 1}^{L}A_{\perp 1}^{*L}) + \{L \leftrightarrow R\}\right), \end{aligned} \tag{C7}$$

$$\begin{aligned} K_{4ss} = \frac{\sqrt{3}}{2}&\left(Im\left(B_{\perp 1}^{L}A_{\parallel 1}^{*L}\right) - Im\left(B_{\parallel 1}^{L}A_{\perp 1}^{*L}\right) + \{L \leftrightarrow R\}\right) \\ &- 2\sqrt{3}\frac{m_l^2}{q^2}\left(Im\left(B_{\perp 1}^{L}A_{\parallel 1}^{*L}\right) - Im\left(B_{\parallel 1}^{L}A_{\perp 1}^{*L}\right) + \{L \leftrightarrow R\}\right), \end{aligned} \tag{C8}$$

$$K_{5s} = \frac{\sqrt{6}}{2}\beta_l\left(Re\left(B_{\perp 1}^{L}A_{\parallel 0}^{*L}\right) - Re\left(B_{\parallel 1}^{L}A_{\perp 0}^{*L}\right) - \{L \leftrightarrow R\}\right), \tag{C9}$$

$$\begin{aligned} K_{5sc} = -\frac{\sqrt{6}}{2}&\left(Re\left(B_{\parallel 1}^{L}A_{\parallel 0}^{*L}\right) - Re(B_{\perp 1}^{L}A_{\perp 0}^{*L}) + \{L \leftrightarrow R\}\right) \\ &+ 2\sqrt{6}\frac{m_l^2}{q^2}\left(Re\left(B_{\parallel 1}^{L}A_{\parallel 0}^{*L}\right) - Re(B_{\perp 1}^{L}A_{\perp 0}^{*L}) + \{L \leftrightarrow R\}\right), \end{aligned} \tag{C10}$$

$$K_{6s} = \frac{\sqrt{6}}{2}\beta_l\left(Im\left(B_{\parallel 1}^{L}A_{\parallel 0}^{*L}\right) - Im(B_{\perp 1}^{L}A_{\perp 0}^{*L}) - \{L \leftrightarrow R\}\right), \tag{C11}$$

$$\begin{aligned} K_{6sc} = -\frac{\sqrt{6}}{2}&\left(Im\left(B_{\perp 1}^{L}A_{\parallel 0}^{*L}\right) - Im\left(B_{\parallel 1}^{L}A_{\perp 0}^{*L}\right) + \{L \leftrightarrow R\}\right) \\ &+ 2\sqrt{6}\frac{m_l^2}{q^2}\left(Im\left(B_{\perp 1}^{L}A_{\parallel 0}^{*L}\right) - Im\left(B_{\parallel 1}^{L}A_{\perp 0}^{*L}\right) + \{L \leftrightarrow R\}\right), \end{aligned} \tag{C12}$$

**Appendix D:**

Input values that are used in the work are recorded in the following table [71].

| Parameter | Values |
|---|---|
| $m_\mu$ | $(105.66 \pm 0.0000024)$ MeV |
| $m_e$ | $(0.51 \pm 0.0000000031)$ MeV |
| $m_\tau$ | $(1776.86 \pm 0.12)$ MeV |
| $\mathcal{B}_{\Lambda^*}$ | $(0.45 \pm 0.01)$ |
| $\alpha_s$ | $(0.1181 \pm 0.0011)$ MeV |
| $\alpha_e$ | $1/127.925\,(16)$ |
| $m_{\Lambda^*}$ | $(1519.54 \pm 0.17)$ MeV |
| $m_{\Lambda_b}$ | $(5619.60 \pm 0.17)$ MeV |
| $\tau_{\Lambda_b}$ | $(1.47 \pm 0.01) \times 10^{-12}$ sec |
| $G_F$ | $(1.166 \pm 0.0000006) \times 10^{-5}$GeV$^{-2}$ |
| $\mu$ | 4.8 GeV |
| $m_b$ | $(4.18 \pm 0.04)$ GeV |
| $\|V_{tb}\|$ | $(1.019 \pm 0.025)$ |
| $\|V_{ts}\|$ | $(39.4 \pm 2.3) \times 10^{-3}$ |